\documentclass[11pt,preprint]{aastex}
\usepackage{lscape}
\usepackage{hyperref}
\begin{document}

\title{GALA cookbook}

\author{Alessio Mucciarelli}
\affil{Dipartimento di Fisica \& Astronomia, Universit\`a 
degli Studi di Bologna, Viale Berti Pichat, 6/2 - 40127
Bologna, ITALY}
\email{alessio.mucciarelli2@unibo.it}

\section{Introduction}

GALA \citep{m13} is a code written in standard Fortran 77 and aimed at finding the 
best atmospheric parameters and the abundance of individual elements 
by using the equivalent widths (EWs) of metallic lines, 
providing graphical and statistical tools to evaluate the goodness of the solution. 
The derivation of the abundances is performed by using a modified version of the WIDTH9 
code\footnote{http://wwwuser.oat.ts.astro.it/castelli/sources/WIDTH.html.} 
(originally 
developed by R. L. Kurucz) in its Linux version \citep{sbordone04}.
In the current release, GALA can manage the classical grid of ATLAS9 models 
computed by \citet{castelli04}, the grid of new ATLAS9 models computed for the 
APOGEE survey \citep{meszaros} and 
the MARCS models with the standard composition \citep{gustaf98}.
When the ATLAS9 models are used, new model atmospheres are calculated 
starting from an existing guess model 
and according to the pre-tabulated Opacity Distribution Functions (ODF) and 
Rosseland opacity tables. When MARCS are used, each new model 
is obtained by interpolating within the MARCS grid.

\section{Installation}

Once you have downloaded the archive file {\tt gala.tar.gz} from the website
\begin{center}
\url{www.cosmic-lab.eu/Cosmic-Lab/Products.html},  
\end{center}
type the commands\\
\\
{\tt gunzip gala.tar.gz}\\
{\tt tar -xvf gala.tar}\\

These commands unpack the archive file creating a directory named  {\tt GALA/}.
Within this directory there are three subdirectories ( {\tt src/}, {\tt bin/} and {\tt tutorial/}) together 
with some information files. The directory {\tt src/} includes all the source files and the 
{\tt Makefile} needed to install GALA. The directory {\tt bin/} includes the executables created 
after the installation. In the directory {\tt tutorial/} there are some examples of the 
input files for GALA as reference.\\ 
\\
(1)~\underline{Install ATLAS9}\\
\\
Before starting the installation of GALA, you need to install in your machine the ATLAS9 
code if you plan to use these models (obviously, if you plan to use only MARCS models 
you can avoid to install ATLAS9). 
GALA includes a dynamical call to the last version of this code, available 
on the website of F. Castelli\footnote{http://wwwuser.oat.ts.astro.it/castelli/sources/atlas9codes.html}. 
Download the source code {\tt atlas9mem.for} with the command\\
\\
{\tt  wget -r -nd http://wwwuser.oat.ts.astro.it/castelli/sources/atlas9/atlas9mem.for}\\
\\
and compile it\\
\\
{\tt ifort -double-size 64 -save -o atlas9mem.exe atlas9mem.for}\\
\\
Note that GALA assumes that you have this code installed and 
that the executable is named {\tt atlas9mem.exe}. Please do not compile 
ATLAS9 with different names.\\ 
\\ 
(2)~\underline{Set the Makefile}\\
\\
Firstly, you need to properly update the {\tt Makefile}. 
GALA can be compiled with the Intel Fortran Compiler, that you can download from 
the Intel website\footnote{http://software.intel.com/en-us/non-commercial-software-development}, 
after the registration.
Also, the graphical output files are created by the program {\tt galaplot} using 
the SuperMongo\footnote{http://www.astro.princeton.edu/$\sim$rhl/sm/} (SM) libraries (namely 
{\tt libplotsub.a}, {\tt libdevices.a} and {\tt libutils.a}, compiled in single precision) and the 
X11 libraries. Set in the first lines of the {\tt Makefile} the path of your Intel Fortran Compiler 
and that of the directory where the SM libraries are located. \\ 
\\
(3)~\underline{Set some useful paths}\\
\\
Now, you need to properly set in 
the {\tt GALA/src/Paths.f} file the paths of some 
directories used by the code:\\ 
(1)~'exe' = it is the path where the executables of GALA are located 
(the subdirectory {\tt bin/});\\ 
(2)~'at9exe' =  it is the path of the directory where the executable 
{\tt atlas9mem.exe} is located;\\ 
(3)~'at9fc' = it is the path where the ATLAS9 models by \citet{castelli04} are saved, see Section.\ref{howmodel};\\ 
(4)~'at9apo' = it is the path where the ATLAS9 models  by \citet{meszaros} are saved, see Section.\ref{howmodel};\\ 
(5)~'marcs' = it is the path where the MARCS models with the standard composition 
by \citet{gustaf98} are saved, see Section.\ref{howmodel}.\\ 
The directories of the ATLAS9 models need a precise structure. In each 
directory you need to create three subdirectories, namely {\tt ODF/}, {\tt MODELS/} and 
{\tt MODELS-new/}, 
the first including the ODFs and the Rosseland opacity 
tables, the second including the grid of model atmospheres used as guess models, while 
the third is the directory where the new models created during the run of GALA will be 
saved. 
The directory for the MARCS models does not need a specific substructure. Examples of 
these subdirectories, as well as more details about the use of the different model grids, 
are discussed in Sect.~\ref{howmodel}.\\
Note that you do not need to have necessarily all these models to use GALA. If you plan to use only 
a given kind of models, you can delete or leave blank 
the paths of the other models in the {\tt Paths.f} file 
(but in this case do not try to call them when you run GALA).\\ 
\\
(4)~\underline{And now...install GALA!}\\
\\
To install GALA, type the command\\
\\
{\tt make all}\\ 
\\
and the executables will be saved in the subdirectory {\tt /bin}.\\ 
Finally, put the path of the subdirectory {\tt bin/} in your 
login file according to the shell environment of your 
machine (for instance in the configuration file .bashrc or .tcshrc).
The directory {\tt /tutorial} includes examples for the three input files. You can 
check the installation of GALA by using these files and simply type {\tt gala}.\\ 
Note that GALA is dimensioned to manage up to 1200
spectral lines for each star (parameter INL in the source file {\tt Declare.f}). 
If a given star has more than this number of measured transitions, GALA skips this star 
providing a warning message. For our experience this value is reasonable but
if you think the this limit is too low for your stars, you can 
increase the parameter INL and recompile the code (note that a uselessly
high value for INL will lead to a waste of computer memory).

\section{Installation problems}
We list some common installation problems (typically related to the combination Fortran compiler/system architecture) 
and their possible solutions. 
\begin{itemize}
\item The SM libraries compiled in double precision can give problems during the compilation or 
(when the compilation well works) {\tt galaplot} creates corrupted or badly drawn postscript files. We recommend to 
use for {\tt galaplot} SM libraries compiled in single precision. Because it is usual to compile 
SM in double precision, we suggest to perform two distinct installations of SM, in order to save wherever you want 
the libraries compiled in single precision. The instructions to perform the installation of SM are explained 
in the {\tt sm.install} file in the directory of the SM source code.
\item A warning message appears during the compilation of {\tt galaplot} like\\ 
\\
{\tt ld: warning in /Users/alessio/MAGAZZINO/SMONGO-lib-ELE/libplotsub.a, file is not of required architecture}\\
\\
In this case, you can try to 
compile {\tt galaplot} telling the compiler to generate the code for a specific architecture.
You need only to add the option '-m32' (if your machine is a 32 architecture) or '-m64' 
(if your machine is a 64 architecture) in the {\tt Makefile}\\
\\
{\tt \$(FORTRAN) -m32 -traceback - galaplot galaplot.f mplot1.f -lX11 -L\$(SMLIB)}\\
{\tt -lplotsub -ldevices -lutils}.\\ 
\item The SM commands under Fortran are unrecognized, with warning messages like\\
\\
{\tt Undefined symbols:
  "\_sm\_limits\_", referenced from:
      \_mplot\_ in ifortLFD1nn.o}\\
\\      
This happens because some incompatibilities between the different names of the 
SM commands as defined by Fortran and C compilers can occur. 
The users with experiences with other tools devoted to spectroscopic analysis 
(as MOOG and DAOSPEC) have already dealt with these problems. 
We refer the reader to the Chapter 4 of the 
{\sl DAOSPEC Cookbook}\footnote{http://www.bo.astro.it/$\sim$pancino/docs/daospec.pdf} as useful 
documentation to solve some of these problems. \\
With the last versions of the Intel Fortran Compiler, the SM commands have 
an extra '\_' as suffix (not needed if you compile {\tt galaplot} with G77 Fortran Compiler for instance). 
The source file {\tt mplot1.f} calls the SM commands as {\sl sm\_limits\_} and so on. 
Note that we provide a modified version of this file (named {\tt mplot2.f}) where 
the SM commands are called with their standard names (i.e. {\sl sm\_limits}); you can 
try to compile {\tt galaplot} with this source file, if the error messages shown above appear.
\item
If you cannot compile {\tt galaplot} with the Intel Fortran Compiler 
because of incompatibilities between the SM libraries and the Fortran commands 
as called in the source code, 
you have two alternative options: {\sl (1})~try to compile only {\tt galaplot} 
with another Fortran compiler (as G77 for instance), substituting the file {\tt mplot1.f} with {\tt mplot2.f}; 
{\sl (2)}~if you cannot 
compile {\tt galaplot} with other compilers, you can perform the installation 
of GALA without {\tt galaplot}. In this case, you can easily disable the graphical option 
by commenting in the file {\tt gala.f} the 
line \\
\\
{\tt call system('galaplot')}\\
\\
and then compile the code with the command 
{\tt make gala}.
\end{itemize}

\section{Executable files}

The executable files that will be created in the subdirectory {\tt bin/} are:
\begin{enumerate}
\item {\tt gala} --- the executable file to run GALA; 
\item {\tt galaplot} --- this program produces the graphical output for each analysed star; 
\item {\tt interpol\_marcstoatlas} --- this is a modified version of the code to interpolate   
the MARCS model atmospheres developed by T. Masseron and available in its original 
version at the website http://marcs.astro.uu.se/software.php. The current version 
writes the output MARCS model in the ATLAS9 standard format; 
\item {\tt marcstoatlas} --- this routine reads an usual MARCS model atmosphere
converting it in the ATLAS9 standard format. This routine is a modification of the 
program (developed by B. Edvardsson) available in the Uppsala website 
(http://marcs.astro.uu.se/software.php) and it can be used 
independently by GALA with the command-line instruction\\
\\
{\tt cat $<$ model-name $>$ / marcstoatlas}\\
\\
(where  $<$ model-name $>$ is the input MARCS model),
storing in the Fortran unit 88 the ATLAS9-like output model; 
\item {\tt width9\_gala} --- this is a modified version of the WIDTH9 code available 
on the website of F. Castelli 
\footnote{http://wwwuser.oat.ts.astro.it/castelli/sources/WIDTH.html}. 
The main differences are the formats of the 
input and output files. 
Some control cards of WIDTH9 \citep[see][]{castelli88} have been 
disabled because not used by GALA;
\item {\tt extract\_kur} --- this program reads the classical linelists of Kurucz/Castelli 
\footnote{http://wwwuser.oat.ts.astro.it/castelli/linelists.html} stored in 
a file named {\tt master\_linelist.tmp}
and a file (named {\tt input.dat}) including 
your linelist (in the first column the wavelength in $\mathring{\rm A}$ and in the second column 
the code of the element in the Kurucz formalism: i.e. 26.00 for FeI and 26.01 for FeII). 
The code matches the two files and writes in the output file ({\tt output.dat}) the 
information needed to the GALA input.
\item {\tt write\_autofl} --- this program writes the basic structure of the input file 
{\tt autofl.param} (described in Section~\ref{autofl}), with the proper name of each keyword.\\
\end{enumerate}


\section{Input files}

\begin{itemize}
\item {\tt autofl.param} --- it includes the 
main configuration parameters used by GALA to perform the analysis 
(described in Section 5.1);
\item {\tt list\_star} --- it is the list of the 
stars to analyse and their guess atmospheric parameters (described in Section 5.2);
\item a file (one for each star listed in {\tt list\_star}) with extension {\tt .in}, 
including the list of the measured lines, their EWs and their 
atomic parameters (described in Section 5.3).
\end{itemize}

\subsection{Layout of {\tt autofl.param}}
\label{autofl}
The input file {\tt autofl.param} includes two groups of keywords, namely the basic 
parameters and the subordinate parameters. 
The command {\tt write\_autofl} writes the basic structure of the file (with some 
parameters already filled by default values); note that the order of the keywords 
(as well as the presence or not of the header lines) does not really matter.
A template of this file is available in the subdirectory {\tt tutorial/}.

\begin{figure}[h]
\epsscale{0.6}
\plotone{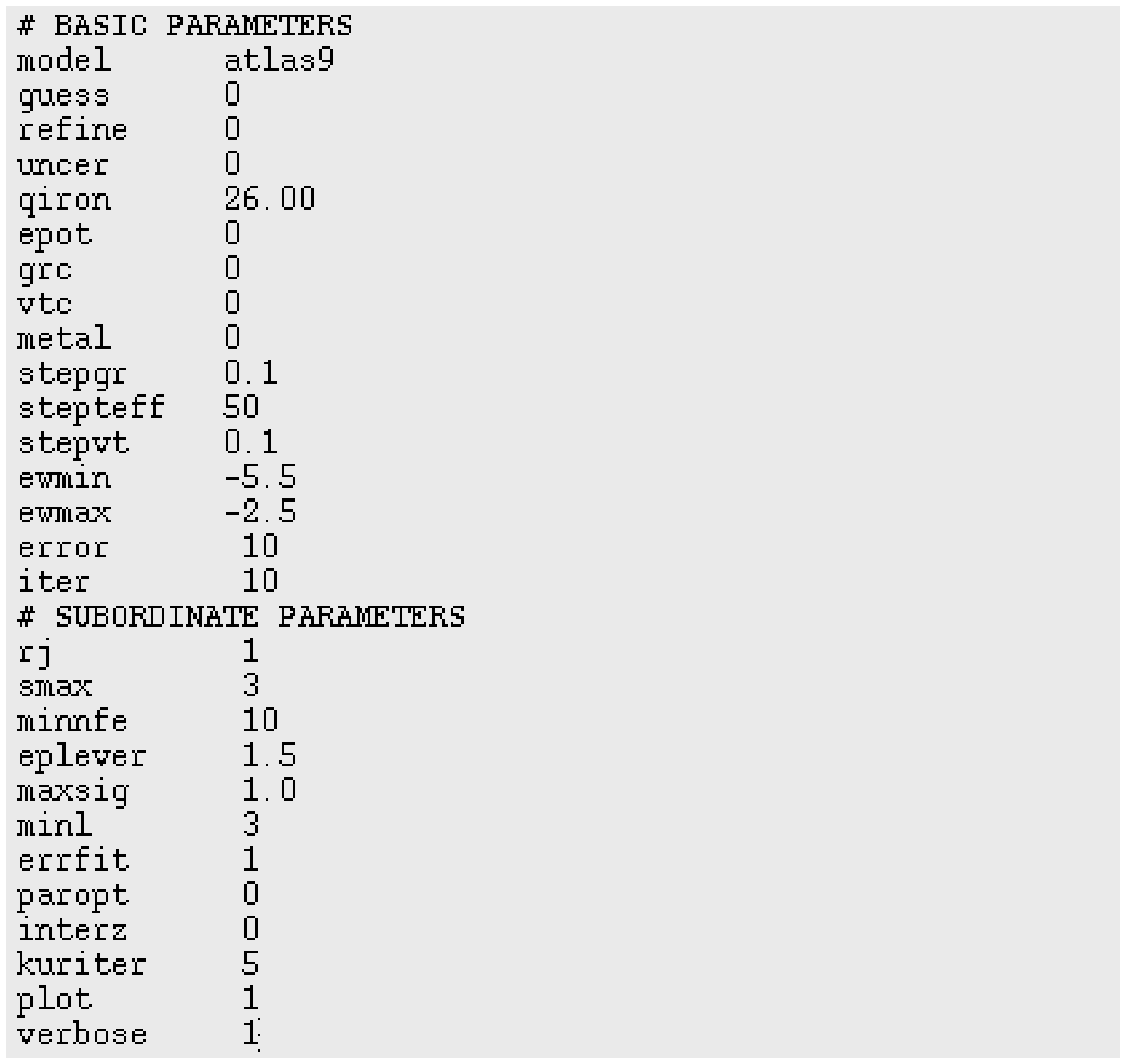}
\end{figure} 

(1) Basic parameters: these are the parameters that you need to 
set in order to perform the analysis: 

\begin{itemize}
\item {\bf model} 
specifies the kind of model atmospheres; the accepted values are \\ 
-{\sl atlas9} for the ATLAS9 model atmospheres available in the website by F. Castelli;\\ 
-{\sl apogee\_cm150} APOGEE-ATLAS9 grid with [C/Fe]=--1.50;\\
-{\sl apogee\_cm125} APOGEE-ATLAS9 grid with [C/Fe]=--1.25;\\
-{\sl apogee\_cm100} APOGEE-ATLAS9 grid with [C/Fe]=--1.00;\\
-{\sl apogee\_cm075} APOGEE-ATLAS9 grid with [C/Fe]=--0.75;\\
-{\sl apogee\_cm050} APOGEE-ATLAS9 grid with [C/Fe]=--0.50;\\
-{\sl apogee\_cm025} APOGEE-ATLAS9 grid with [C/Fe]=--0.25;\\
-{\sl apogee\_cp000} APOGEE-ATLAS9 grid with [C/Fe]=+0.00;\\
-{\sl apogee\_cp025} APOGEE-ATLAS9 grid with [C/Fe]=+0.25;\\
-{\sl apogee\_cp050} APOGEE-ATLAS9 grid with [C/Fe]=+0.50;\\
-{\sl apogee\_cp075} APOGEE-ATLAS9 grid with [C/Fe]=+0.75;\\
-{\sl apogee\_cp100} APOGEE-ATLAS9 grid with [C/Fe]=+1.00;\\
-{\sl marcs} for the MARCS model atmospheres with the standard chemical composition.\\

We refer to Section~\ref{howmodel} to detail about the download of these models and the 
way to manage them in order to use correctly GALA.
Alternatively, you can specify 
the name of a given model (in ATLAS9 format)
\footnote{If you wish to use a model atmosphere calculated with the ATLAS12 code 
\citep{castelli05a}, you need to erase the 22 lines corresponding to the 'ABUNDANCE TABLE' 
control cards. Also, to use a specific MARCS model, you have to put it in ATLAS9 format 
with the program {\tt marcstoatlas}.} located in the working directory. 
If the specified model is found in the current directory, all the optimization 
options (together with the guess and refinement working-blocks) are switched off 
and only the analysis working-block is executed (with fixed parameters). Otherwise, 
if the specified model does not exist in the directory, GALA stops the procedure and 
exits with a warning message.

\item {\bf guess} 
enables the guess working-block, before the analysis working-block. 
The parameter is the number of iterations performed during this working-block. 
Guess=~0 (or negative values) disables the execution of this working-block.
If you decide to use this option, we suggest to adopt at least 3 iterations 
for the guess working-block.

\item {\bf refine} 
enables the refinement working-block, after the analysis working-block. 
The parameter accepts the values 0 (disabling the working-block) and 1 
(enabling the working-block).

\item {\bf uncer} 
enables the calculation of the uncertainties in the derived chemical abundances 
due to the atmospheric parameters. 
The accepted values are \\
0 : no uncertainties calculation is performed;\\  
1 : the uncertainties are calculated according to the prescriptions by 
\citet{cayrel04}, only if $T_{\rm eff}$ has been optimized spectroscopically;\\ 
2 : together with the uncertainties estimated with the previous method, also the 
uncertainties due to the atmospheric parameters following the classical method 
are calculated, varying each time one parameter only (keeping the others fixed) 
and without taking into account the covariance terms.

\item {\bf qiron} 
indicates the code of the element (with its ionization stage) whose lines are 
used to perform the optimization. The formalism is the same used in the Kurucz codes 
(i.e. 26.00 for Fe~I and 26.01 for Fe~II). In principle, GALA accepts all the elements 
but obviously only a few elements (i.e. Ti or Ni) have a sufficient number of available 
lines to perform a reliable spectroscopic optimization.

\item {\bf epot} 
enables the spectroscopic optimization of the temperature. 
The accepted values are \\
0 : the temperature is not optimized and rather fixed to the input value;\\
1 : the temperature is found by erasing any trend in the 
A(El)\footnote{A(El) indicates for a given element El the 
number abundance as calculated by WIDTH9.}---$\chi_{ex}$ plane (where 
$\chi_{ex}$ is the excitation potential in eV). 

\item {\bf grc} 
enables the spectroscopic optimization 
of the gravity.\\
0 : the gravity is fixed to the input value; \\
1 : gravity is optimized in order to obtain the same abundance 
from neutral and single ionized iron lines; \\ 
2 : if this option is chosen, GALA reads in 
{\tt list\_star} the term $\epsilon={\rm log(4GM\pi\sigma/L)}$ and for each iteration 
log~g is computed as log~g=$\epsilon$+4$\cdot T_{\rm eff}$ (see Section 5.2);\\
3 : the gravity is computed with 
a second degree relation logg=~A+B$\cdot T_{\rm eff}$+C$\cdot T_{\rm eff}^2$. 
In this case the coefficients A, B and C are provided in the {\tt list\_star} file 
(see Section 5.2).

\item {\bf vtc} enables the optimization of the 
microturbulent velocity. \\ 
0 : the velocity is fixed to the input value;\\ 
1 : the microturbulent velocity is found by erasing any trend in the A(El)---EWR
plane, where EWR is defined as $\log(EW/\lambda)$, where EW and 
$\lambda$ are expressed in the same units. 

\item {\bf metal} enables the optimization of the overall metallicity of the 
model atmosphere, that is chosen in order to reproduce the average abundance 
derived from lines of the element specified by the parameter {\tt qiron}
(within the metallicity step of the employed models grid).

\item {\bf stepgr} the step used to investigate log~g in the analysis and 
refinement working-blocks;

\item {\bf stepteff} the step used to investigate $T_{\rm eff}$ in the analysis and 
refinement working-blocks;

\item {\bf stepvt} the step used to investigate $v_{turb}$ in the analysis and 
refinement working-blocks;

\item {\bf ewmin, ewmax} these are the minimum and maximum value for 
EWR. All lines outside the EWR range defined by these parameters 
are excluded from the analysis.
If you do not know a priori the optimal EWR range, 
a first run of GALA with a wide range including all lines is 
recommended (with EWR ranging from {\tt ewmin}=~-10 to {\tt ewmax}=~0), 
followed by a more informed second run, with EWR limits deducted from visual 
inspection of the graphical outputs.

\item {\bf error} 
the maximum error in percentage in EW, in the case in which the error 
for each line is provided in the linelist file. All the lines with error larger than this 
threshold are excluded from the analysis.

\item {\bf iter} the maximum number of allowed iterations. 
This parameter avoids the risk of infinite loops.
A reasonable value is 10 (if the input parameters are not so far from the 
real solution), but you should adapt this value to her/his particular needs.

\end{itemize}

(2) Subordinate parameters: these parameters specify 
some finer details of the analysis; we suggest to use the  
default values for these parameters, unless you have specific needs\\

\begin{itemize}

\item {\bf rj} specifies whether the line-rejection uses 
the mean ({\tt rj}=0) or the median value (default {\tt rj}=~1).

\item {\bf smax} is the number of $\sigma$ used in the line-rejection process 
(default {\tt smax}=~3). 

\item {\bf minnfe} corresponds to the minimum number of surviving lines 
that allows to perform the optimization. In principle, the spectroscopic 
optimization of the atmospheric parameters is a statistical method which requires a
large (at least 40-50) set of lines. Optimizations derived with a low number of lines 
are affected by fluctuations due to the small number statistics and the lines distribution
in the A(El)---$\chi_{ex}$ and A(El)---EWR planes (but these uncertainties are taken into account 
in the calculation of the errors through the Jackknife bootstrap technique). 
You should be careful that a very low number of lines (typically ten or less) would provide 
very dangerous results
(default {\tt minnfe}=~10).

\item {\bf eplever} 
the spectroscopic derivation of $T_{\rm eff}$ needs a large range of $\chi_{ex}$ covered by the 
used lines, in our opinion at least 1.5--2 eV. For smaller range, the 
optimization of $T_{\rm eff}$ can be unreliable. This keyword specifies the minimum 
$\chi_{ex}$ range (in eV) allowed to perform the optimization of $T_{\rm eff}$ (if requested). 
If the range of $\chi_{ex}$ (calculated considering 
only the lines surviving line-rejection) 
is smaller than {\tt eplever}, then the optimization of $T_{\rm eff}$ is swichted off; eventually, if the 
other parameters are to be optimized, their optmization continues
(default {\tt eplever}=~1.5).

\item {\bf maxsig} is the maximum dispersion by the mean or median for the element chosen with {\tt qiron} 
allowed to run GALA. When the dispersion exceeds this value, all the optimizations are disabled 
and the analysis is performed with the input atmospheric parameters specified in the {\tt list\_star} file
(default {\tt maxsig}=~1.0).

\item {\bf minl} corresponds to the minimum number of lines to perform the 
process of line-rejection. If the number of lines is less than the value of {\tt minl} 
($N_{lines}\le$minl),
the lines are rejected only on the basis of uncertainty in EWs ($\sigma_{EW}$) and of the pair of values 
{\tt ewmin}--{\tt ewmax} (default {\tt minl}=~3).

\item {\bf errfit} specifies the kind of linear fit performed on the 
A(El)---$\chi_{ex}$ and A(El)---EWR planes. \\ 
0 : the linear fits are computed without considering the uncertainties in x and/or y;\\
1 : the linear fits are computed by taking into account the uncertainties in A(El);\\
2 : the linear fit in A(El)---$\chi_{ex}$ is computed including the uncertainties in A(El) 
and that in A(El)---EWR including the uncertainties both in EWR and A(El), following the method 
implemented by \citet{press}.

The use of the uncertainties in the slopes computation is recommended, but only if you have 
reliable errors in EWs, able to provide a realistic ranking among the used lines. If you 
do not trust your EW uncertainties, use {\tt errfit}=~0.
Note that if {\tt errfit}=~1 or 2 but at least one of the input lines has $\sigma_{EW}$=~0.0, the 
inclusion of the errors in the slope calculation is disabled (default {\tt errfit}=~1).

\item {\bf paropt} specifies the kind of optimization parameter used for 
$T_{\rm eff}$ and $v_t$:\\
0 : optimizes these parameters by using the slopes (default);\\ 
1 : by using the Spearman ranking coefficient.

\item {\bf interz} enables the interpolation to the zero value of the 
optimized parameters ({\tt interz}=~1).
This option is included to provide visually nicer results if one wants to plot temperatures 
and gravities in a $T_{\rm eff}$---logg diagram. It interpolates the slope in the A(El)--$\chi_{ex}$ 
and the A(El)I-A(El)II to zero, and determines the corresponding $T_{\rm eff}$ and logg, 
thus avoiding the grid effect in the results. 
If this interpolation is switched off, all results will be in steps specified by 
the {\tt stepgr}, {\tt stepteff}, and {\tt stepvt} parameters, while if the interpolation is switched on, 
they will be distributed more continuously. Note, however, that all uncertainties will still be 
computed on the best slopes, not on the interpolated zero slopes (default {\tt interz}=~0) . 

\item {\bf kuriter} specifies the number of blocks of 15 iterations each used 
by ATLAS9 to calculate a new model atmosphere (default {\tt kuriter}=~5). 
This option is ignored if the keyword {\tt model} is referred to MARCS models.

\item {\bf plot} enables the graphic output in postscript format\\
0 : does not produce a plot;\\ 
1 : saves the plot of the results in a postscript file (default option).

\item {\bf verbose} specifies the verbosity level on the terminal\\
0: no message at all. It advises only when GALA ends the analysis;\\
1: the sequence of the analysed stars is shown, without additional information;\\ 
2: this is like {\tt verbose}=~1 but also specifies the main steps performed 
by GALA (without additional information about them);\\
3: this is the default verbosity level. It is like {\tt verbose}=~2 but also displays 
information about the line rejection, the derived abundances and 
concerning the Guess Working-Block.
\end{itemize}

\subsection{Starlist file}
The file {\tt list\_star} includes the sequence of the stars that you 
plan to analyse, together with the input atmospheric parameters. 
The structure of the file depends basically by the kind of the adopted optimization  
for the surface gravity (as specified by the basic parameter 'grc' in the file 
{\tt autofl.param}).
\begin{itemize}
\item
if you adopted grc=~0 or 1, each line of the file will be as follows:\\

\begin{figure}[h]
\plotone{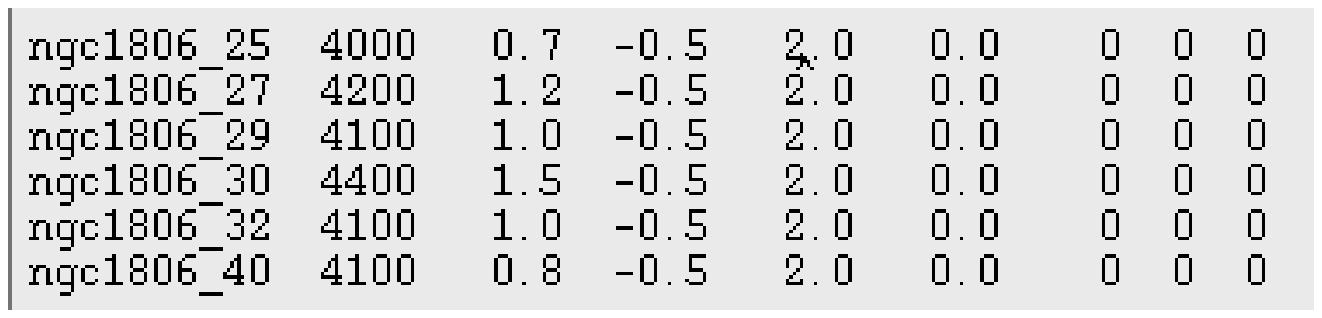}
\caption{Example of the {\tt list\_star} file when the option grc=0 or 1 is selected.
}
\label{inp1}
\end{figure}

\begin{enumerate}
\item the first column contains the root name of the input files, without the extension {\tt .in};\\ 
\item the four following columns contain the guess atmospheric parameters in the 
following order: $T_{\rm eff}$, log~g, [M/H] and $v_t$.
Note that if you provide a wrong value of [M/H] (not available in the chosen grid of models), 
GALA changes it and assumes the closest metallicity;
\\ 
\item the sixth column contains the value of the $\alpha$-enhancement 
adopted in the computation of the ATLAS9 models. Also in this case, if you provide a 
value of [$\alpha$/Fe] not available among the models, the closest value is used.
If you use MARCS models 
GALA neglects this parameter;\\ 
\item the last three columns are the errors associated to $T_{\rm eff}$, logg and $v_t$ respectively. 
If the parameters are optimized spectroscopically, these uncertainties 
are ignored (you can also put 0 for all of them, as in the shown case); 
on the other hand, if a given parameter 
is fixed and the uncertainties with the classical method  are requested 
({\tt uncer}=2 in {\tt autofl.param}), these values are used to vary each parameter 
computing again the abundances. If in the input file the parameter uncertainties are 
0 but you request to calculate the abundance uncertainties, 
the steps used to investigate the parameter space (and defined by the keywords 
{\tt stepteff}, {\tt stepgr} and {\tt stepvt} among the basic parameters in {\tt autofl.param}) 
are assumed as parameter errors.
\end{enumerate}

\item if you adopted grc=~2, the input file will be\\

\begin{figure}[h]
\plotone{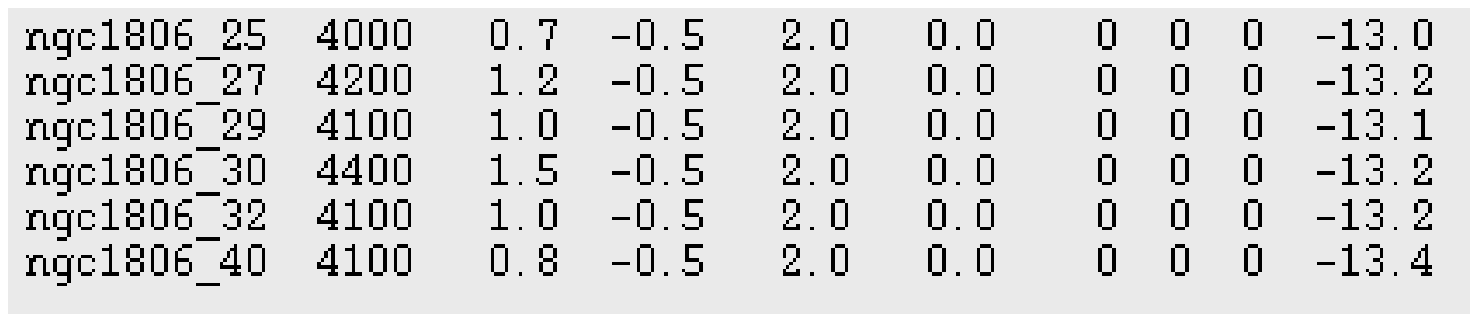}
\caption{Example of the {\tt list\_star}  file when the option grc=2 is selected.}
\end{figure}

The additional tenth column $\epsilon$ is calculated according to the Stefan-Boltzmann equation as 
$\epsilon={\rm log(4GM\pi\sigma/L)}$, where G is the gravitational constant, 
$\sigma$ the Boltzmann constant and M and L are the mass and the luminosity of the star. 
In this case the surface gravity will be calculated as logg=~$\epsilon + 4\cdot\log{T_{\rm eff}}$

\item if you adopted grc=~3, the input file will be\\

\begin{figure}[h]
\plotone{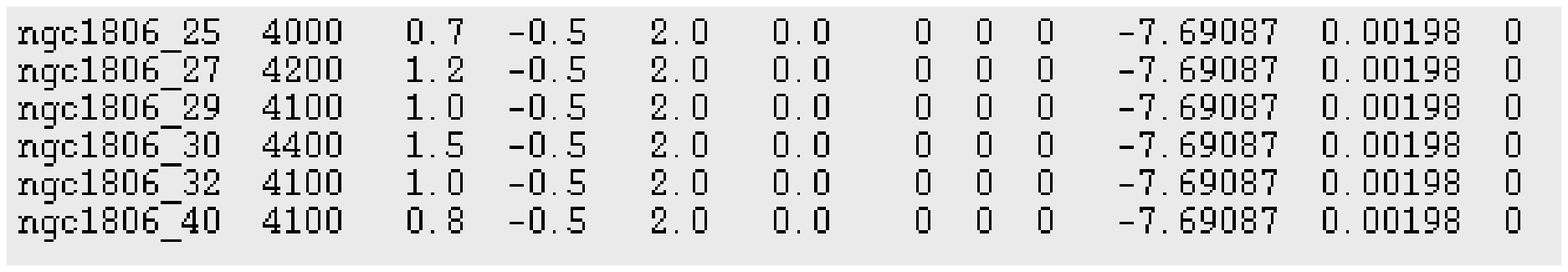}
\caption{Example of the  {\tt list\_star}  file when the option grc=3 is selected.}
\end{figure}

where A, B and C are the coefficients of the quadratic relation 
log~g=~A+B$\cdot T_{\rm eff}$+C$\cdot T_{\rm eff}^2$.

\end{itemize}

\subsection{Linelist file}
For each star (specified by the $<$rootfilename$>$ in the {\tt star\_list} file) 
an input file {\tt $<$rootfilename$>$.in} is needed. If the linelist file of a given star listed 
in {\tt list\_star} is not available in the directory, GALA provides a warning 
message and moves to the next star.
Fig.~\ref{inp1} shows an example of a linelist input file for GALA.
Note that this file does not need a specific numeric format, but it must 
contain the following information for each measured line:
\begin{itemize}
\item the wavelength (expressed in $\mathring{\rm A}$);
\item the observed EW (expressed in m$\mathring{\rm A}$);
\item the uncertainty in EW (expressed in m$\mathring{\rm A}$); 
\item the code of the element in the usual Kurucz notation;
\item the logarithm of the oscillator strength;
\item the excitation potential (expressed in eV);
\item the logarithm of the radiative  damping constant, $\gamma_{rad}$
\footnote{When one of the $\gamma$ constants is 0, WIDTH9 calculates its value
adopting some theoretical relations \citep[see][for more details]{castelli05b}.}; 
\item the logarithm of the Stark damping constant, $\gamma_{stark}$; 
\item the logarithm of the Van der Waals  damping constant, $\gamma_{VdW}$;
\item the velocity parameter $\alpha$ as defined by \citet{barklem00} and 
usually available for many metallic lines both in the Kurucz/Castelli linelist 
and VALD databases. If the parameter $\alpha$ is not available for a given line, 
the value can be substituted with 0.
\end{itemize}

\begin{figure}[h]
\plotone{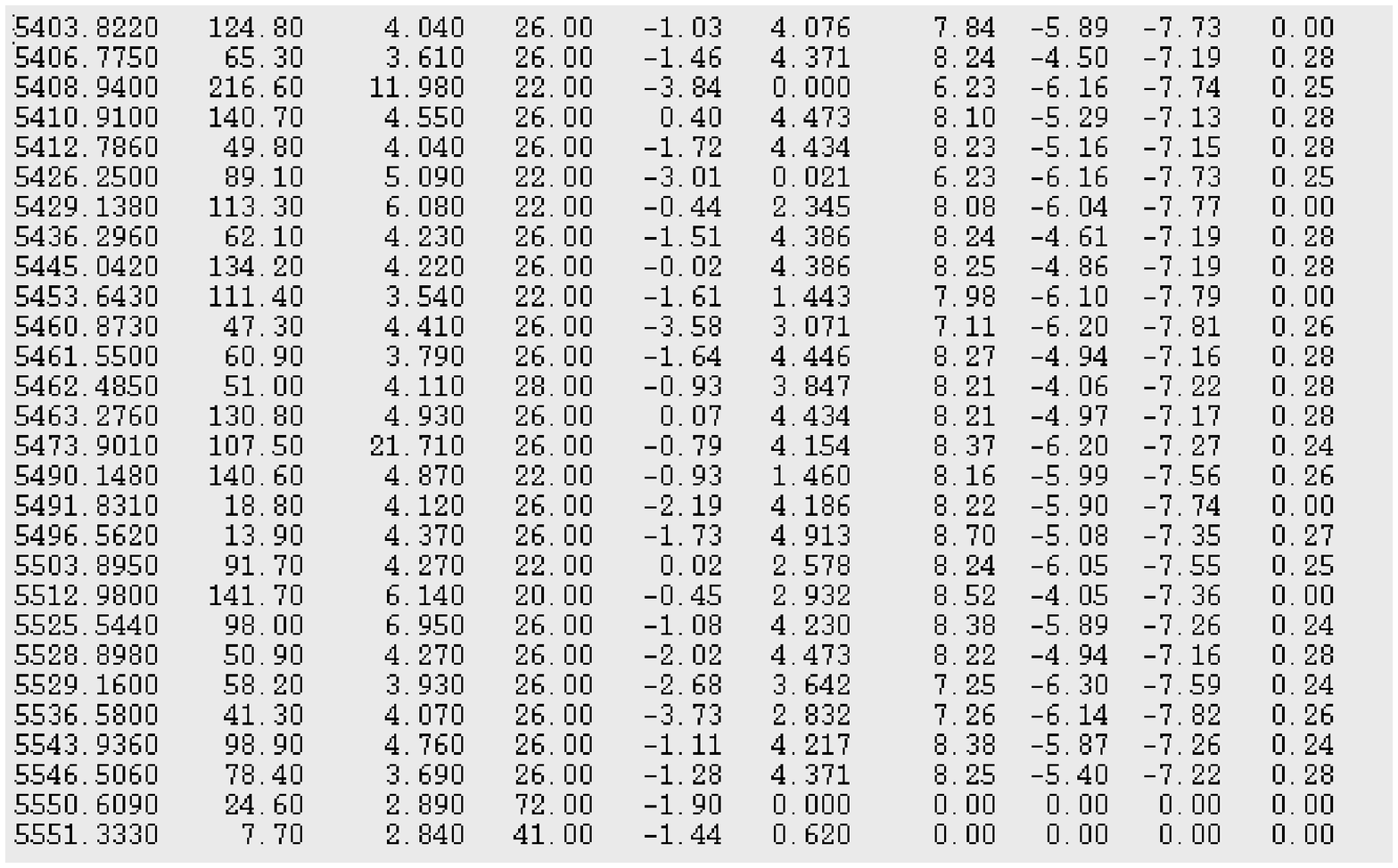}
\caption{Example of the linelist input file ({\tt $<$rootfilename$>$.in}) for GALA.
}
\label{inp1}
\end{figure}

\section{Output files}

For each star listed in the file {\tt list\_star}, GALA will create 
three output files.

\begin{figure}[h]
\epsscale{1.0}
\plotone{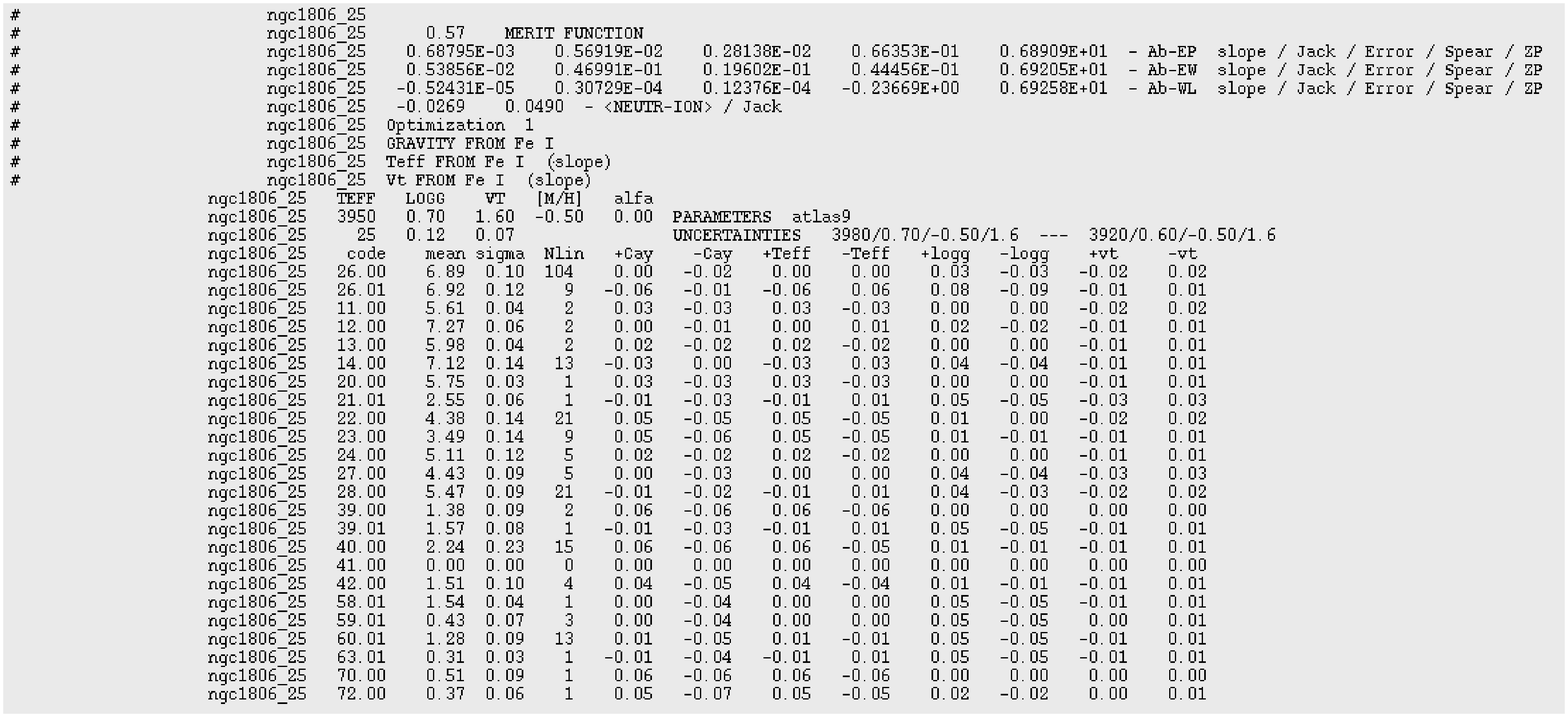}
\caption{Example of the output file ({\tt $<$rootfilename$>$.abu}) including information about 
atmospheric parameters, uncertainties and average abundances.
}
\label{abu}
\end{figure}

(1)~ {\tt $<$rootfilename$>$.abu} summarizes the derived parameters, the uncertainties and the mean 
abundance for each element available in the input file {\tt $<$rootfilename$>$.in}.
Fig.~\ref{abu} shows an example of this kind of output file. 
The first thirteen lines are referred to the adopted analysis:
\begin{itemize}
\item the second line is the value of the merit function, that provides 
an estimate of the goodness of the solution; \\
\item third, fourth and fifth lines concern the correlation among the lines 
in the A(El)--EWR , A(El)---$\chi_{ex}$ and A(El)---$\lambda$ planes, 
summarysing the slope, the Jackknife uncertainty in the slope, the formal error in the slope, 
the Spearman rank correlation coefficient and the zero-point of the linear fit;\\
\item the sixth line is the difference between the average abundances of the chosen element 
derived from neutral and single ionized lines, together with its Jackknife uncertainty;\\
\item the seventh line is a flag ({\tt Optimization}) about the adopted procedure: 'Optimization 1' 
indicates that the analysis has been performed as specified in {\tt autofl.param}, while 
'Optimization 0' advises that during the analysis the optimization of one or more 
parameters has been switched off. 
The following three lines specify as log~g, $T_{\rm eff}$ and $v_t$ have been 
optmized. The {\tt Optimization} flag is useful to identify rapidly stars for which the 
analysis has had some problems or failures.\\
\item The final atmospheric parameters are written in the twelfth line, together 
with the adopted model atmosphere (i.e. 'atlas9', 'marcs' or the name of the 
used individual model). In case of ATLAS9 models, if some atmospheric layers of 
the final model atmosphere do not converge
\citep[according to the convergence criteria proposed by ][]{castelli88}, 
the number of unconverged layers is specified.
Also, the label {\tt PARAMETERS} allows to easily extract the final parameters 
(for instance with {\tt grep} or {\tt awk} commands).\\
\item The uncertainties in the parameters are written in the thirteenth  line; at the end of the line, 
the parameters of the model atmospheres obtained with the procedure described by \citet{cayrel04} 
are reported (only if the option {\tt uncer}=~1 or 2 is used and $T_{\rm eff}$ has been spectroscopically 
optimized).
Note that these uncertainties in the atmospheric parameters are calculated by using the Jackknife 
technique for the optimized parameters: when a parameter is fixed, the uncertainty is that 
provided by you in {\tt list\_star}, if the latter is 0, the step of the grid 
is taken as representative of the error.\\
\item The following lines provide for each element the mean abundance, the dispersion of the mean, the 
number of used lines, the net variation of the abundance by using the two models obtained with the 
\citet{cayrel04} method (columns '+Cay' and '-Cay') and the net variation of the abundance by varying 
each time one only parameter (columns '+Teff', '-Teff', '+logg', '-logg', '+vt', '-vt').
Additionally, a summary file (named {\tt abundance.final}) is provided, including 
all the {\tt $<$rootfilename$>$.abu} files for the stars listed in 
{\tt list\_star}; 
\end{itemize}

\begin{figure}[h]
\epsscale{1.0}
\plotone{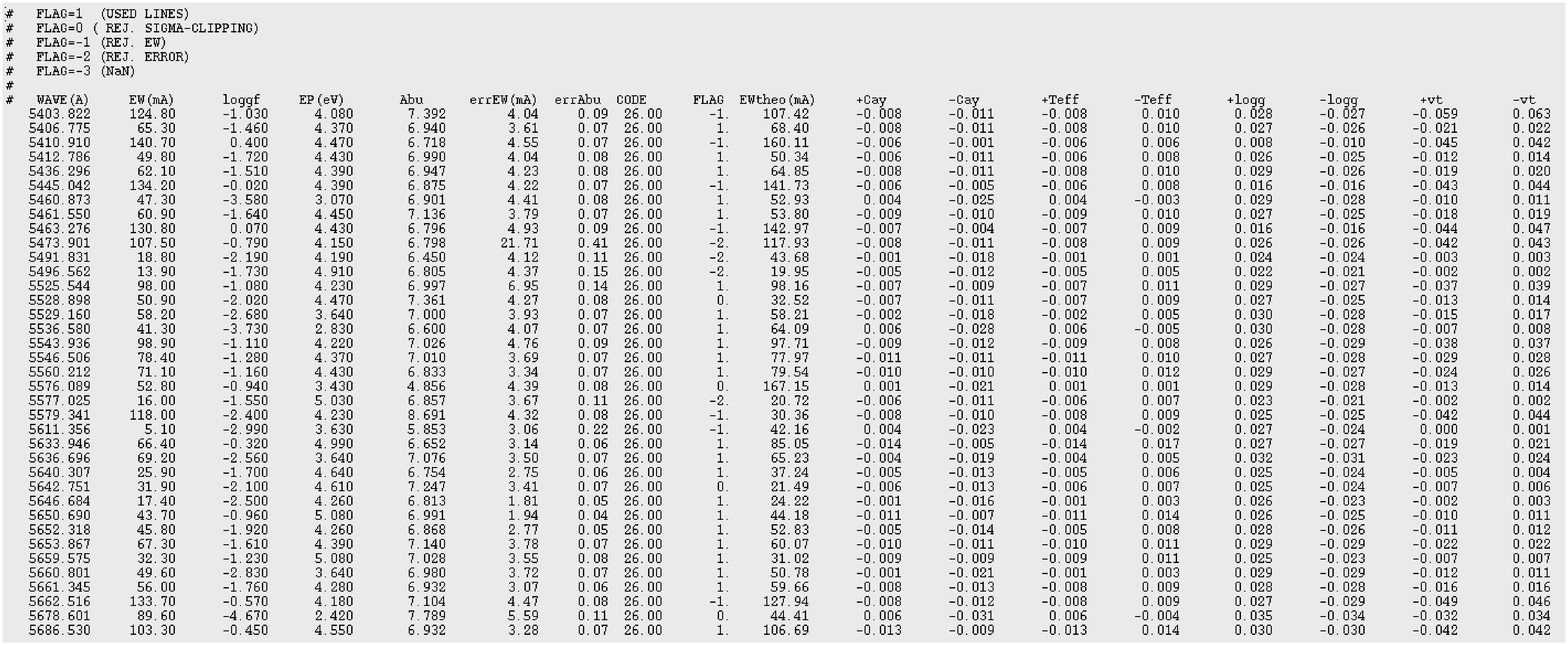}
\caption{Example of the output file ({\tt $<$rootfilename$>$.EW}) including information for each 
individual line.
}
\label{ew}
\end{figure}

(2)~ {\tt $<$rootfilename$>$.EW} includes the main information for each transition 
included in the input file {\tt $<$rootfilename$>$.in}. Fig.~\ref{ew} shows a part of this file as reference.
The first rows of the file explain the meaning of the flag about 
the line rejection. Briefly:\\ 
flag = 1 the line is used;\\ 
flag = 0 the line is rejected according to the $\sigma$-clipping
algorithm \citep[see][]{m13};\\ 
flag = --1 the line is rejected because outside the range of valid EWR 
values specified by 'ewmin' and 'ewmax' in 'autofl.param';\\ 
flag = --2 the line is rejected because its error is larger than the 
boundary value 'error' in {\tt autofl.param};\\ 
flag = --3 if the abundance calculation performed by WIDTH9 does not 
converge, producing a 'NaN', the value is substituted with 99.999 and 
the line excluded by the analysis.\\
The rest of the file includes for each transition:
\begin{itemize}
\item the wavelength (expressed in $\mathring{\rm A}$);
\item the observed EW (expressed in m$\mathring{\rm A}$);
\item the logarithm of the oscillator strength;
\item the excitation potential (expressed in eV);
\item the derived abundance;
\item the error of EW, if provided in input (expressed in m$\mathring{\rm A}$) ;
\item the error in abundance  obtained by varying the observed EW  
of +1$\sigma_{EW}$;
\item the code of the element in the usual Kurucz notation;
\item the flag related to the line rejection;
\item the theoretical EW (expressed in m$\mathring{\rm A}$) calculated by WIDTH9
through Gaussian integration of the line profile by using the nominal 
metallicity of the best model \citep[see][for details]{castelli05b};
\item the net abundance variation calculated according to the method described by 
\citet{cayrel04} and keeping fixed the effective temperature at $T_{\rm eff}+\sigma T_{\rm eff}$
and $T_{\rm eff}-\sigma T_{\rm eff}$;
\item the net abundance variations due to the variation of each parameter 
($\pm\sigma T_{\rm eff}$,$\pm\sigma$log~g, $\pm\sigma v_{t}$).
\end{itemize}

\begin{figure}
\epsscale{1.0}
\plotone{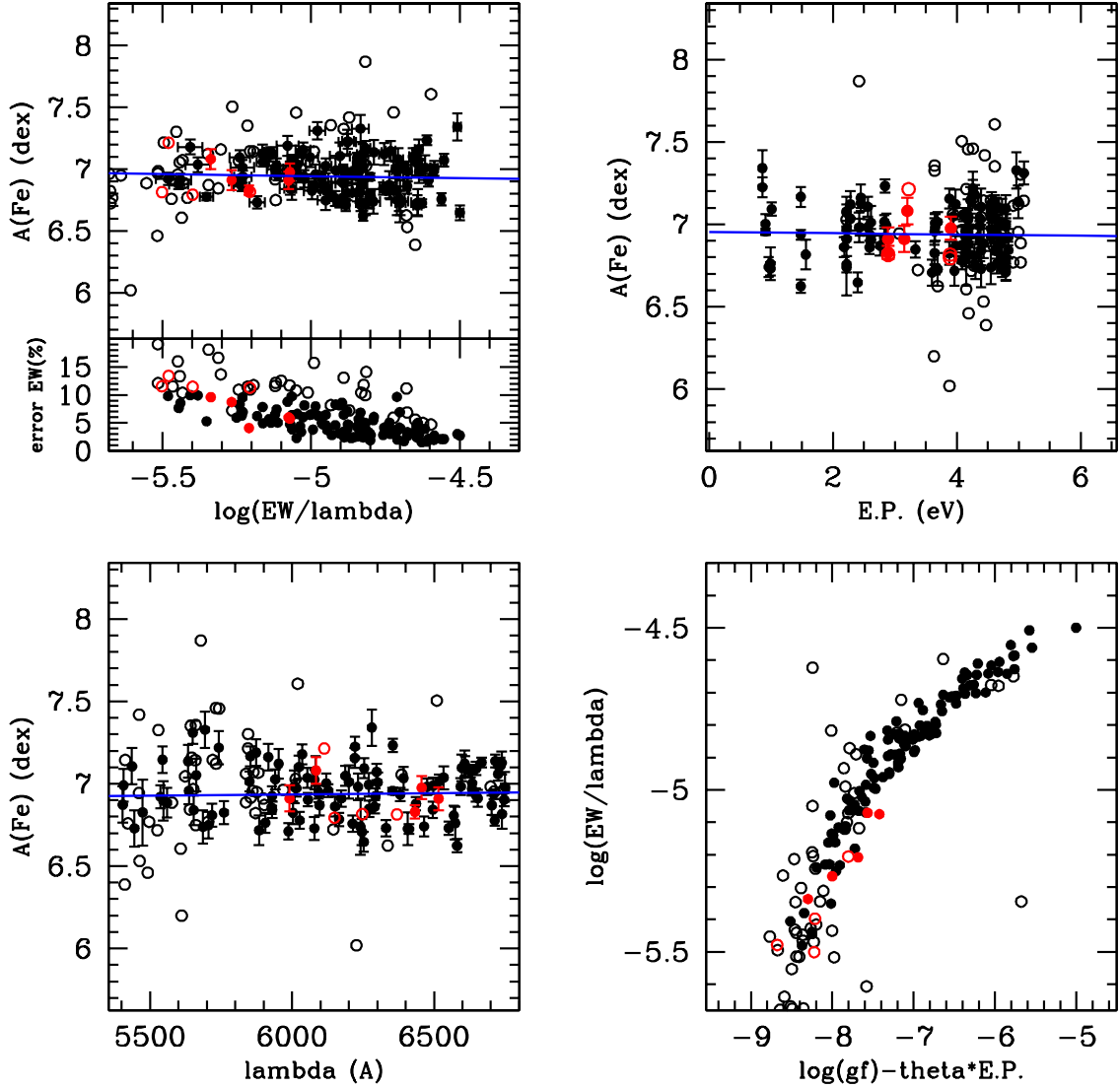}
\caption{Example of the graphical output ({\tt $<$rootfilename$>$.ps}) of GALA. 
Left-upper panel shows the behaviour of the abundances as a function of the 
reduced EW, EWR (while the small panel shows the behaviour of the EW error 
as a function of EWR). 
Right-upper and left-lower panels plot the behaviour of the abundances as a function of $\chi_{ex}$ 
and of the wavelength, respectively. 
The right-lower panel shows an empirical curve of growth.
In all the panels the filled circles are the used lines and empty circles are the 
rejected lines. Black symbols are the lines used in the optimization of $T_{\rm eff}$ and 
$v_t$ (Fe~I in this case), while red symbols are the lines in the other ionization stage 
(Fe~II in this case). Blue lines are the linear  best-fits.
}
\label{ps}
\end{figure}

(3)~ {\tt $<$rootfilename$>$.ps} is the graphical output of GALA where the the main results are 
shown and an example of this output is shown in  Fig.~\ref{ps}. 
The first three panels show the behaviour of the abundances for the chosen element 
as a function of EWR, $\chi_{ex}$ and $\lambda$;
the small lower panel in the first panel shows the behaviour of the EW error as a function of EWR.
The blue lines are the linear fits obtained in each plane.
The last panel shows the curve of growth obtained by plotting EWR as a function of 
the theoretical EW, defined as EWT=~loggf - $\theta\chi_{ex}$.\\ 
Black points are the lines in the ionization stage used to optmize $T_{\rm eff}$ and $v_t$, 
while red points the lines of the same species but in the other ionization stage; for instance, 
if you specify 26.00 (Fe~I) in the 'qiron' parameter, black points in the graphical 
output will be the neutral iron lines and the red points the Fe~II lines. Empty points are the 
lines rejected during the analysis (the reason of the rejection is specified in the {\tt $<$rootfilename$>$.EW} file).

\section{How to manage the grids of model atmospheres}
\label{howmodel}
The automatization of the chemical analysis as implemented by GALA 
needs the management of wide grids of model atmospheres, in order to 
freely explore the parameter space. 
Here we briefly explain how you can obtain the grids of models.
It is worth to recall that you do not need to download all the 
models described in the following: you can only have the grid of models 
that you will use for your analysis. The only caveat is that you cannot 
call a kind of model in the {\tt autofl.param} file if you do not have 
a directory with all these models.

If you decide to adopt ATLAS9 models for your analysis, you need to create 
in the corresponding directory three sub-directories, named ODF/, MODELS/ and 
MODELS-new/ (the latter is empty and it will be filled with new models 
progressively using GALA).

Note that the capability to calculate a given ATLAS9 model atmosphere 
depends on the availability of the ODF computed with the requested chemical composition. 
The input guess model atmosphere has to be as close as possible to the new model in terms 
of $T_{\rm eff}$, log~g and metallicity in order to allow a good convergence,
but in principle, the output model is independent from the input one.
Thus, when a new ATLAS9 model is requested during the analysis process, GALA looks for 
the ODF and Rosseland table corresponding to the requested metallicity, if these tables do exist, 
the closest model with that chemical composition is picked as input guess model (GALA recognizes automatically the 
closest model),
otherwise GALA skips the star and advises you with a warning message that the 
selected metallicity is incompatible with the metallicities encoded in GALA.
The extension (in terms of metallicity, $\alpha$-enhancement and microturbulent velocity) 
of each ATLAS9 grid is written in the source code of GALA, allowing an easy implementation 
of additional ATLAS9 grids in the next releases of GALA.


{\bf ATLAS9 (F.Castelli)}\\
This dataset includes ODFs calculated for 5 values of microturbulent velocities 
(namely, 0, 1, 2, 4 and 8 km/s) and two levels of $\alpha$-enhancement, [$\alpha$/Fe]=0.0 and +0.4. 
GALA uses the metallicities between --5.5 and +0.5 dex with a step of 0.5. 
To easily obtain all the needed files, we provide a {\tt wget}-based procedure. 
Move the code {\tt downl\_fx.csh} in the directory specified in the file {\tt GALA/src/Paths.f} 
and type\\
\\
{\tt ./downl\_fc.csh}\\
\\
However, in the following we describe the correct way to obtain and manage the ATLAS9 files.
\begin{enumerate}
\item In the directory specified in the file {\tt GALA/src/Paths.f} for this grid of models, 
create three subdirectory, named {\tt ODF}, {\tt MODELS} and {\tt MODELS-new}.
In the first subdirectory, put ODFs and Rosseland tables with [M/H] between --5.5 and +0.5 dex 
(at step of 0.5) and with both the values of $\alpha$-enhancement.
You can download the ODFs from the website of 
F. Castelli
\begin{center}
http://wwwuser.oat.ts.astro.it/castelli/odfnew.html
\end{center} 
(you need 
only the {\sl big} ODFs, do not download the {\sl lit} ODFs), 
as well as the tables with the Rosseland mass absorption 
coefficients
\begin{center}
http://wwwuser.oat.ts.astro.it/castelli/kaprossnew.html.
\end{center}

\item 
The grids of models are available in the same website
\begin{center}
http://wwwuser.oat.ts.astro.it/castelli/grids.html.
\end{center} 

For the metallicities used by GALA, all the models 
of a given [M/H] are available in a single ASCII file 
(for instance, master file for [M/H]=+0.5 and [$\alpha$/Fe]=~0.0 
is named {\tt ap05k2odfnew.dat}, while for [M/H]=+0.5 and [$\alpha$/Fe]=~+0.4 
{\tt ap05ak2odfnew.dat}.) 
Save these (14) files in the subdirectory {\tt MODELS}.
We recommend to download these files (without renaming them), because GALA looks for these 
files to pick the input guess model
(please, avoid downloading the grids with peculiar assumptions, as He enhancements and 
different mixing length values, because GALA does not (yet) use these grids to search for the input model).
Note that when you ask GALA to use a metallicity outside the range --2.5/+0.5, the 
star will be analysed by adopting the closest model.

\item The new ATLAS9 models calculated during the runs of GALA will be saved in the
subdirectory {\tt MODELS-new}, together with a file including the main information 
about the calculation of the model. The new models will be named following the same  
formalism used by \citet{castelli04} for the individual models; for instance, a model 
with $T_{\rm eff}$=~4120 K, log~g=~1.32, $v_{turb}$=2 km/s, [M/H]=--0.5 and [$\alpha$/Fe]=~+0.4 
will be saved as {\sl am05at4120g132k2odfnew.dat}, will its summary file will be named 
{\sl am05at4120g132k2.log\_at9}.
\end{enumerate}


{\bf ATLAS9 (APOGEE)}\\
Recently, \citet{meszaros} computed a new set of ATLAS9 model atmospheres, 
ODFs and Rosseland opacity tables, spanning a huge range of metallicity, 
$\alpha$-enhancement and C abundances. 
In order to avoid the introduction of another input parameter (namely the 
C abundance, [C/Fe]), GALA uses subgrids with fixed C abundances. 
The models (as well as the ODFs and the Rosseland opacity tables) are freely available 
from the website 
\begin{center}
http://www.iac.es/proyecto/ATLAS-APOGEE/
\end{center}
Note that if you plan to use only the models with a given [C/Fe] value, you can easily download 
only the desired files or download all the models and ODFs and then delete these not of interest 
(in order save disk space). In following we describe how to obtain all the models but the same 
instructions are valid also to manage ODFs and models for a given [C/Fe] value.

\begin{enumerate}

\item Also for the ATLAS9-APOGEE grid, create in the corresponding directory 
three subdirectories named {\tt ODF}, {\tt MODELS} and {\tt MODELS-new}. 
On the webpage, the available data are provided in {\tt tar.bz2} archive files 
according to the metallicity. Download the ODFs archive files.

\item The archive file including all the models (and named {\tt all\_mod.tar}) 
is available on the main page of the ATLAS9-APOGEE project. Download this file 
in your subdirectory {\tt MODELS/} and unpack it. The archive includes 
for each chemical composition a subdirectory with all the individual models and 
a master file grouping together all the models of that chemical composition. 
Delete all these subdirectories and leave only the master files.
These files are named according to the chemical composition: for instance, {\tt mm05cm03op00.mod} 
is the master file including all the models with [M/H]=--0.5, [C/Fe]=-0.25 and [$\alpha$/Fe]=+0.0).
Do not rename neither the models nor the ODFs, otherwise GALA will crash because it will be 
not able to find the needed files to calculate a model atmosphere.

\item The new ATLAS9 models will be saved in the subdirectory {\tt MODELS-new}. 
Also for the ATLAS9-APOGEE models, new models and summary files will be named 
according to the nomenclature used to label the original model of the grids. 
Also for the models, do not rename the files. 
For instance, the model with 
$T_{\rm eff}$=~4120 K, log~g=~1.32, $v_{turb}$=2 km/s, [M/H]=--0.5, [C/Fe]=-0.25 and [$\alpha$/Fe]=~+0.0 
will be named {\tt amm05cm03op00t4120g132v20.mod}. 

\end{enumerate}  
  
\begin{figure}[h]
\epsscale{1.0}
\plotone{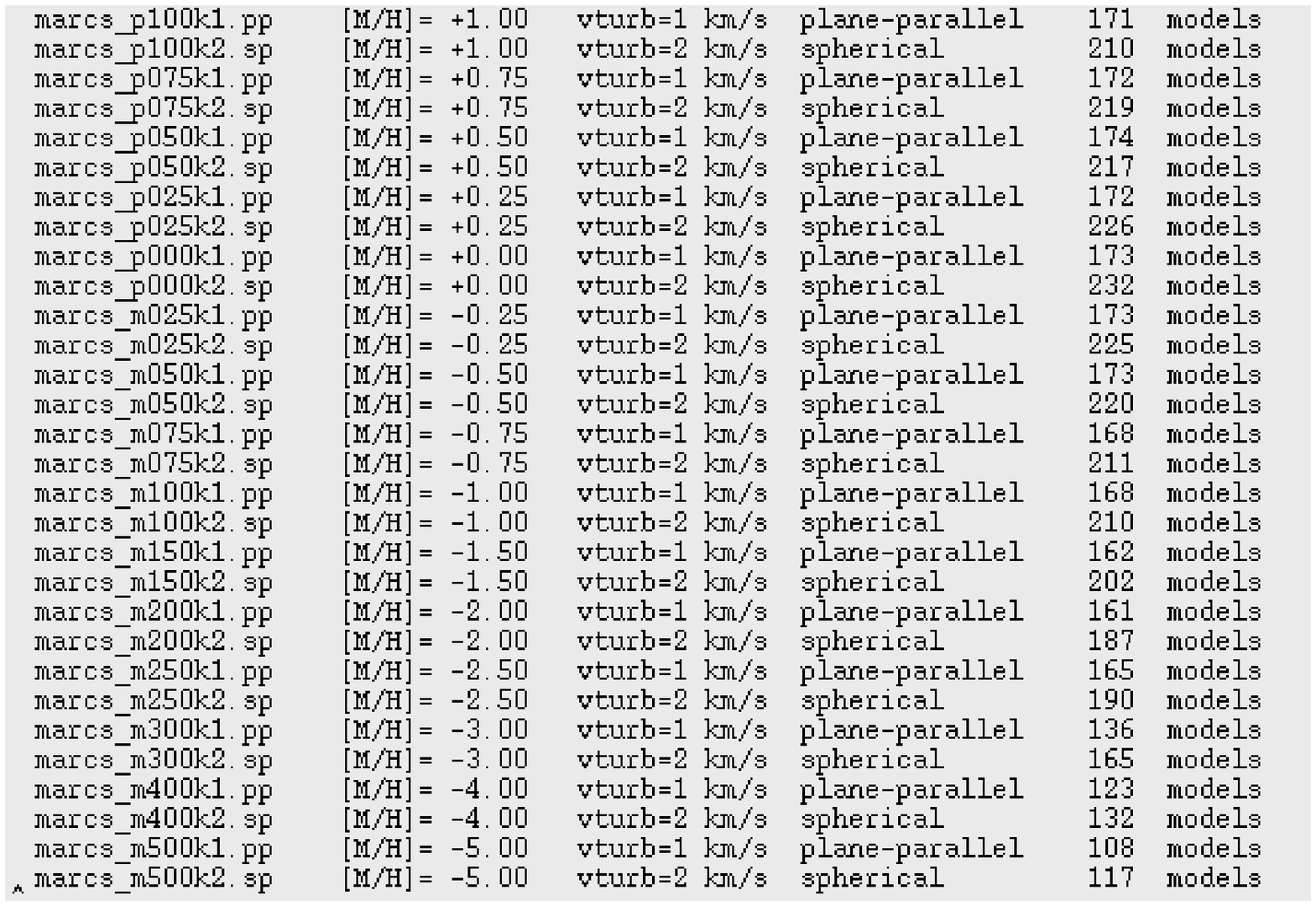}
\caption{List of the files with the MARCS models grouped according to their metallicity, 
microturbulent velocities and geometry.
}
\label{marx}
\end{figure}

{\bf MARCS}\\
MARCS model atmospheres are available at the website 
\begin{center}
http://marcs.astro.uu.se/ 
\end{center}
but unfortunately the entire grid of models cannot be downloaded as a whole, because 
the web-interface allows to download a limited (less than an hundred) number of models at a time. 
If you wish to use the MARCS models, he/she needs to patiently download all the models 
from the website and then group them according to their metallicity, microturbulent velocity (as done for the 
ATLAS9 grids) and geometry (plane-parallel or spherical). GALA is designed to manage only the grids with
standard chemical composition and microturbulent velocities 1 and 2 km/s, but 
modified versions of the code can be provided on request for other sub-grids of MARCS models.
Because the grid of models with standard chemical composition naturally varies the $\alpha$-enhancement 
according to the metallicity, for these models it is not necessary to group them for the value of the 
$\alpha$-enhancement, at variance with the ATLAS9 models. 
Each master file needs to have a precise name. In Fig.~\ref{marx} we summarize the name of these master files, 
as expected by GALA, together with the corresponding metallicity and geometry, and the total number 
of included models.

\end{document}